\documentclass[sigconf]{acmart}

\usepackage{algorithm}
\usepackage{balance}
\usepackage{graphicx}
\usepackage{textcomp}
\usepackage{verbatim}
\usepackage{multirow} 
\usepackage{makecell} 
\usepackage{xspace}
\usepackage{acronym}
\usepackage{subcaption}
\usepackage{cleveref}
\usepackage{pifont}     
\usepackage{xcolor}
\usepackage{framed}
\usepackage{hyperref}

\newcommand{\vllm}[0]{\textsc{LLM4FP}\xspace}
\newcommand{\direct}[0]{\textsc{Direct-Prompt}\xspace}
\newcommand{\grammar}[0]{\textsc{Grammar-Guided}\xspace}
\newcommand{\vari}[0]{\textsc{Varity}\xspace}
\newcommand{\gcc}[0]{\texttt{gcc}\xspace}
\newcommand{\clang}[0]{\texttt{clang}\xspace}
\newcommand{\nvcc}[0]{\texttt{nvcc}\xspace}
\newcommand{\codebleu}[0]{CodeBLEU\xspace}
\newcommand{\real}[0]{\textit{Real}\xspace}
\newcommand{\zero}[0]{\textit{Zero}\xspace}
\newcommand{\posinf}[0]{\textit{+Inf}\xspace}
\newcommand{\neginf}[0]{\textit{-Inf}\xspace}
\newcommand{\nan}[0]{\textit{NaN}\xspace}
\newenvironment{mybox}[1]
  {\MakeFramed{\hsize\linewidth\advance\hsize-\width\FrameRestore}\noindent\textbf{#1}\par}
  {\endMakeFramed}

\crefname{section}{Section}{Sections}
\crefname{table}{Table}{Tables}
\crefname{figure}{Figure}{Figures}
\crefname{subfigure}{Figure}{Figures}
\crefname{definition}{Definition}{Definitions}
\crefname{equation}{Equation}{Equations}
\crefname{example}{Ex.}{Examples}
\crefname{algorithm}{Algorithm}{Algorithms}
\crefname{line}{Line}{Lines}

\AtBeginDocument{%
  }

\copyrightyear{2025}
\acmYear{2025}
\setcopyright{cc}
\setcctype{by}
\acmConference[SC Workshops '25]{Workshops of the International Conference for High Performance Computing, Networking, Storage and Analysis}{November 16--21, 2025}{St Louis, MO, USA}
\acmBooktitle{Workshops of the International Conference for High Performance Computing, Networking, Storage and Analysis (SC Workshops '25), November 16--21, 2025, St Louis, MO, USA}
\acmDOI{10.1145/3731599.3767362}
\acmISBN{979-8-4007-1871-7/2025/11}

\begin{document}

\title{\vllm: LLM-Based Program Generation for Triggering Floating-Point Inconsistencies Across Compilers}

\author{Yutong Wang}
\affiliation{
  \institution{University of California, Davis}
  \country{United States}
}
\email{ytwwang@ucdavis.edu}

\author{Cindy Rubio-Gonz\'alez}
\affiliation{
  \institution{University of California, Davis}
  \country{United States}
}
\email{crubio@ucdavis.edu}

\begin{abstract}

  Floating-point inconsistencies across compilers can undermine the
  reliability of numerical software. We present \vllm, the first
  framework that uses Large Language Models (LLMs) to generate
  floating-point programs specifically designed to trigger such
  inconsistencies. \vllm combines Grammar-Based Generation and
  Feedback-Based Mutation to produce diverse and valid programs. We
  evaluate \vllm across multiple compilers and optimization levels,
  measuring inconsistency rate, time cost, and program
  diversity. \vllm detects nearly 2.5$\times$ the number of
  inconsistencies as the state-of-the-art tool \vari. Notably, most
  of the inconsistencies involve real-valued differences, rather than
  extreme values like \nan or infinities. \vllm also uncovers
  inconsistencies across a wider range of optimization levels, and
  finds the most mismatches between host and device compilers. These
  results show that LLM-guided program generation improves the
  detection of numerical inconsistencies. In practice, numerical
  software and HPC developers can use \vllm to compare compilers and
  select those that provide more accurate and consistent
  floating-point behavior, while compiler developers can use it to
  identify and address subtle consistency issues in their
  implementations.

\end{abstract}


\maketitle

\section{Introduction}
\label{sec:intro}

Floating-point arithmetic is widely used in numerical software,
scientific computing, and scientific simulations. However, reasoning
about its behavior remains challenging. Small rounding errors,
precision issues and different compiler optimizations can all lead to
unexpected behaviors \cite{DBLP:journals/csur/Goldberg91}. For
example, when the same program is compiled with different compilers
and/or executed on different architectures such as CPUs and GPUs,
developers may observe inconsistent results, since different compilers
may generate different code for the same program
\cite{DBLP:conf/dagstuhl/Markstein08}. Such compiler-induced
inconsistencies can lead to numerical bugs in widely used
libraries~\cite{DBLP:conf/kbse/FrancoGR17}, and the floating-point
research community has highlighted the difficulty of ensuring
numerical accuracy and reproducibility across heterogeneous computing
systems~\cite{DBLP:conf/correctness/GopalakrishnanL21}, with
correctness in scientific computing recognized as a major challenge in
High Performance Computing
(HPC)~\cite{DBLP:journals/corr/abs-2312-15640}.

Despite their importance, finding numerical inconsistencies is
challenging. First, the space of possible floating-point programs is
huge, which is difficult to cover all the cases using manual or random
testing. Second, such inconsistencies often happen under certain
compiler optimization levels and combinations of floating-point
operations which are hard to trigger. Third, numerical inconsistencies
are usually small and do not cause a program crash, making it hard to
notice. Therefore, there is a strong need for effective and efficient
approaches to generate floating-point programs that help to discover
numerical inconsistencies across compilers.

To the best of our knowledge, \vari~\cite{DBLP:conf/ipps/Laguna20} is
the state-of-the-art randomized differential testing framework for
generating floating-point programs that trigger numerical
inconsistencies across compilers and/or optimization levels. However,
\vari mainly relies on random program generation and does not make use
of any domain knowledge or prior test results. Many generated programs
fail to trigger any inconsistency, leading to low testing efficiency.
This limits \vari's ability to find more subtle compiler-specific
numerical inconsistencies within a reasonable budget of test programs.
In a parallel line of work, Large Language Models (LLMs) have recently
shown abilities in generating programs that have valid syntax and rich
semantics \cite{DBLP:journals/corr/abs-2406-00515}. Several tools have
begun using LLMs to generate test programs for compiler fuzzing
\cite{DBLP:conf/icse/XiaPTP024, DBLP:journals/pacmpl/YangDLY0J024,
DBLP:journals/fgcs/MunleyJC24, DBLP:conf/sigsoft/Gu23}. However, they
mainly target general-purpose code generation, instead of focusing on
floating-point programs. As a result, existing LLM-based testing
approaches are not well-suited for exposing numerical inconsistencies.

\begin{figure*}[!th]
    \includegraphics[width=0.9\linewidth]{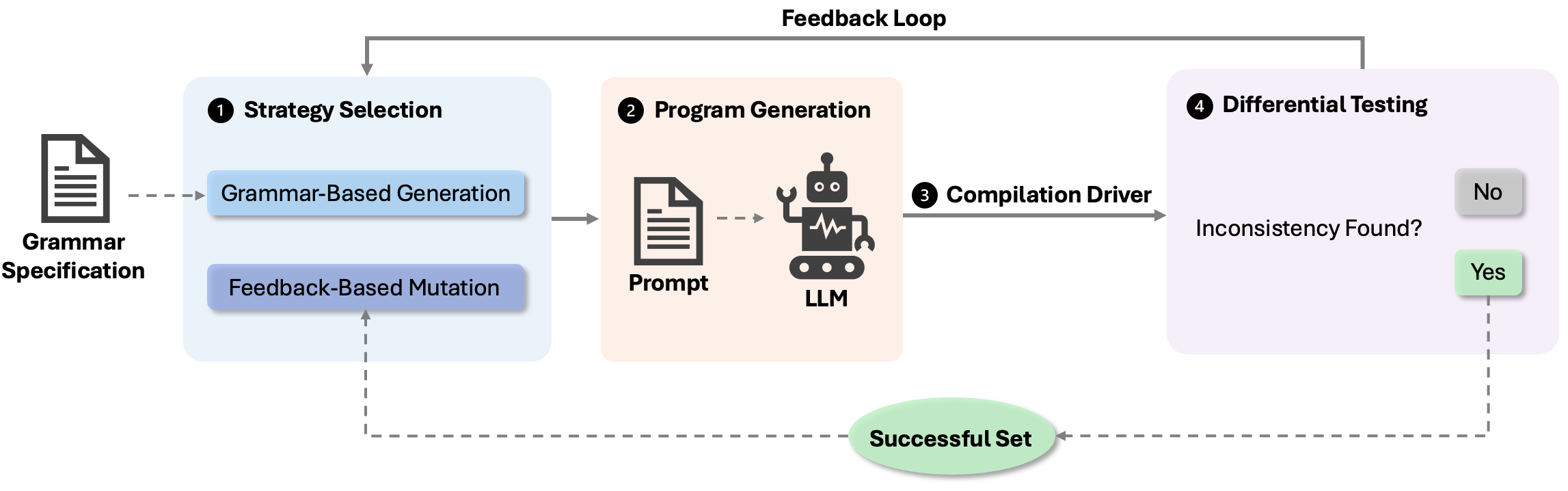}
    \caption{\vllm Overview}
    \label{fig:overview}
\end{figure*}

In this paper, we present \vllm, the first LLM-based framework to
generate floating-point programs specifically designed to trigger
numerical inconsistencies across compilers and/or optimization levels.
Our insight to use LLMs is that they implicitly capture rich prior
domain knowledge from a vast amount of source code ``seen'' during
training. This benefits LLMs to generate programs with meaningful
floating-point operations and diverse code patterns, which are often
missed by existing random generators, and have the potential of
exposing more compiler inconsistencies. 

To realize the potential of using LLMs for floating-point program
generation, we propose two complementary program generation strategies
in \vllm, i.e., Grammar-Based Generation and Feedback-Based Mutation.
The first strategy aims to explore the broad space of floating-point
programs by generating new, well-formed programs from scratch without
relying on previous examples. It gives us control over the structure
of the generated code, ensuring that the programs are syntactically
valid and semantically meaningful. However, we observe that without
feedback, this strategy may lead the LLM to generate similar program
patterns which limits the diversity. In contrast, the second strategy
leverages feedback by mutating previously successful programs that
triggered numerical inconsistencies. By showing the LLM an effective
program and instructing it to generate a different one, we encourage
it to explore novel yet promising programs.

We evaluate \vllm across multiple compilers and optimization levels.
Compared with the state-of-the-art tool
\vari~\cite{DBLP:conf/ipps/Laguna20} and two LLM-based baselines, our
approach consistently detects more floating-point inconsistencies
while generating diverse test programs. Among all approaches, \vllm
achieves the highest inconsistency rate of 29.33\%, roughly 2.5$\times$ that of \vari, 11.93\%. Concretely, \vllm discovers 5,280
inconsistencies out of 18,000 total pairwise comparisons across
compilers and optimization levels. These improvements highlight
\vllm's ability to uncover subtle numerical issues that are often
missed by existing methods. Furthermore, \vllm achieves the highest program diversity among all approaches. Notably,
most of the inconsistencies involve normal and subnormal numerical
values instead of extreme values such as \nan or infinities, which
indicates that \vllm not only ﬁnds more inconsistencies, but also
exposes subtler numerical divergences that are harder to detect than
obvious extreme-value errors.

To summarize, this paper makes the following contributions:

\begin{itemize}
    \item We propose \vllm, the first approach that uses LLMs to
    generate floating-point programs specifically designed to trigger
    numerical inconsistencies across compilers.

    \item We design two complementary generation strategies, i.e.,
    Grammar-Based Generation which helps LLMs produce well-formed
    floating-point programs from scratch, and Feedback-Based Mutation
    which reuses prior successful examples to explore diverse and
    effective program patterns.

    \item We conduct a comprehensive evaluation across multiple
    compilers and optimization levels. \vllm outperforms existing
    approaches in both inconsistency detection rate and program diversity.

\end{itemize}
\section{Technical Approach}

\subsection{Overview}
\cref{fig:overview} shows the overall workflow of our approach, \vllm,
for discovering floating-point inconsistencies across compilers. The
first stage is to select a strategy for generating candidate programs.
We propose two complementary strategies: (1) Grammar-Based Generation,
which uses a given grammar specification to guide the structure of the
generated programs; and (2) Feedback-Based Mutation, which randomly
chooses a previously generated program that triggered numerical
inconsistencies across compilers as basis to generate more programs.
Once the prompt is constructed, we invoke the LLM to generate a new
floating-point C program. The next stage is the Compilation Driver,
which prepares the program for execution on both CPUs and GPUs. Thus
this stage handles C to CUDA code translation  and compilation using a
configurable set of compilers and optimization flags. Finally,
Differential Testing executes the programs on both host and device to
detect inconsistencies in the results. If the generated program
triggers any numerical inconsistencies, then it is added into the
``successful'' program set for potential reuse in the Feedback-Based
Mutation phase. This creates a feedback loop in which successful
programs help guide the LLM to generate other inconsistency-triggering
programs.

\subsection{Floating-Point Program Structure}
\label{sec:program-structure}

Instead of directly prompting the LLM to generate numerical programs
without any constraints, we specify (1)~a high-level program structure
to facilitate differential testing, and (2)~a syntax grammar that
specifies the internal structure of the numerical code to reduce the
probability of generating invalid programs. Our approach uses the
LLM's flexibility while guiding it to generate floating-point programs
with a consistent structure tailored for exposing numerical
inconsistencies. \vllm adopts the high-level program structure and
grammar first introduced in \vari~\cite{DBLP:conf/ipps/Laguna20},
which has been shown to be simple yet effective in guiding the
generation of floating-point programs that expose numerical
inconsistencies. The above also simplifies the comparison between
\vllm and \vari in our evaluation in \cref{sec:eval}.

\paragraph{High-Level Structure} The generated program must contain
only two functions: \emph{main} and \emph{compute}. The \emph{compute}
function takes scalar floating-point arguments, performs a sequence of
arithmetic operations, and computes a scalar floating-point number as
result, which is printed to standard output. The latter assumption
helps when performing differential testing. This function is called
from \emph{main}. 

\paragraph{Internal Structure} The internal structure of the function
\emph{compute} is determined by a grammar, shown in
\cref{fig:grammar}. The grammar guides the LLM to produce meaningful
floating-point computations by capturing common patterns from HPC
programs~\cite{DBLP:conf/ipps/Laguna20}. These patterns are likely to
influence how compilers generate and optimize code. Specifically, the
grammar specification allows to indicate the floating-point precision
level, either single or double precision, to use in the program. It
also includes arithmetic expressions involving basic operators in
$\{+, -, *, /\}$, parentheses ``$()$'' for grouping, and invoking
functions from the standard C Math Library. The grammar also allows
nested loops to reflect typical HPC control structures, as well as
conditional branches. Additionally, the use of temporary
floating-point variables is possible, which may store either scalar
values or arrays. Index-based array accesses are allowed. For example,
\vllm can generate programs that contain code patterns such as
\texttt{for (i=0; i<N; i++) a[i] = ...;}. Nested loops with array
indexing are also supported.

\begin{figure}[t]
\begin{verbatim}
<function> ::= "void" "compute" 
"(" <param-list> ")" "{" <block> "}" 

<param-list> ::= <param-declaration>
| <param-list> "," <param-declaration>  

<param-declaration> ::= "int" <id> 
| <fp-type> <id> 
| <fp-type> "*" <id>  

<assignment> ::= "comp" <assign-op> <expression> ";" 
| <fp-type> <id> <assign-op> <expression> ";"  

<expression> ::= <term> 
| "(" <expression> ")" 
| <expression> <op> <expression>  

<term> = <identifier> | <fp-numeral>

<block> ::= {<assignment>}+ 
| <if-block> <block> 
| <for-loop-block> <block>  

<if-block> ::= "if" "(" <bool-expression> ")" 
"{" <block> "}"  

<for-loop-block> ::= "for" "(" <loop-header> ")" 
"{" <block> "}"  

<bool-expression> ::= <id > <bool-op> <expression>  

<loop-header> ::= "int" <id> ";" <id> "<" <int-numeral> 
";" "++" <id>
\end{verbatim}
\caption{Grammar Specification for \texttt{compute} Function}
\label{fig:grammar}
\end{figure}

\subsection{LLM-Guided Program Generation}

We design two strategies to guide the LLM in generating floating-point
programs: Grammar-Based Generation and Feedback-Based Mutation. The
first program generated is always produced using the Grammar-Based
strategy, since we assume that no examples are provided to start with.
Once programs that trigger inconsistencies have been found, we
randomly choose between the Grammar-Based Generation and the
Feedback-Based Mutation strategies for each new generation, based on a
predefined probability.

\subsubsection{Grammar-Based Program Generation}

This strategy aims to guide the LLM to generate well-formed
floating-point programs from scratch, without relying on prior
examples. We carefully design the prompt to ensure that the generated
floating-point programs are valid and useful for exposing numerical
inconsistencies.

The prompt begins with a general instruction to create a random but
valid floating-point C program. Next, the instruction includes:
(1)~the precision, double or single, to use for floating-point
variables, (2)~the program structure, including both high-level and
internal structures, as described in \cref{sec:program-structure}, and
(3)~guidelines to improve robustness and code quality. Finally, the
prompt instructs the LLM to output plain code only, with no formatting
or explanation.

Specifically for the guidelines, we restrict the library usage to
\texttt{stdio.h}, \texttt{stdlib.h}, and \texttt{math.h}. Without 
this constraint, the LLM may introduce unsupported or missing 
headers, leading to compilation failures. We also explicitly require 
all variables to be initialized, and instruct the LLM to avoid
undefined behavior. Our approach integrates these good coding
practices directly into the prompt to reduce the number of invalid
programs and improve the reliability of inconsistency detection.

\subsubsection{Feedback-Based Mutation}
This strategy uses previously generated programs that have triggered
numerical inconsistencies to create new and diverse variants. By
showing the LLM an example of a successful program and explicitly
instructing it to generate a different one, we encourage it to avoid
repeating the same structure while still learning which patterns are
effective. Our experiments confirm that this strategy helps discover
more compiler inconsistencies, as well as achieves better program
diversity.

The prompt first asks the LLM to change a given floating-point C
program to create a new one that \emph{behaves} differently. Next, the
instruction includes: (1)~the precision, double or single, to use for
floating-point variables, (2)~the high-level program structure, as
described in \cref{sec:program-structure}, (3)~guidelines to improve
robustness and code quality, similar to those in the Grammar-Based
strategy, (4)~a list of mutation strategies to consider, including
reordering or deeply nesting arithmetic expressions, changing numeric
constants, introducing new control flow like nested loops or
conditionals, using different Math Library functions, and inserting
intermediate computations, (5)~a concrete previously successful
example which is randomly selected from the set of successful programs
that triggered inconsistencies. Finally, we include the
output formatting instruction at the end of the prompt as in the last
strategy. In this way, the prompt instructs the LLM to preserve key
aspects of the original program, including its high-level structure,
floating-point precision, and successful computational patterns, while
generating novel variations.

\subsection{Finding Floating-Point Inconsistencies}
\label{subsec:finding-incons}

After generating a program, in the Compilation Driver stage, we prepare 
the generated program for execution on both CPUs and GPUs. Following the 
approach used in \vari, we convert the generated C program into CUDA 
by defining the \emph{compute} function as a \texttt{\_\_global\_\_} 
kernel, which is launched from the \emph{main} function using a single 
block and a single thread. We then compile the C version using multiple 
host compilers, and the CUDA version using a GPU compiler, each under 
various optimization levels. Only binaries that compile successfully 
are passed to the next stage.

In the last stage, we perform differential testing~\cite{DBLP:journals/dtj/McKeeman98}
to check whether different compiler configurations produce inconsistent 
outputs for the same program and inputs. Specifically, for each program, 
we compare the results from the standard output for every pair of 
compilers under each optimization level. A \emph{floating-point inconsistency} is 
recorded when two outputs are not equal in their bitwise representations, 
i.e., the hexadecimal encoding of the floating-point result, such as when 
two 64-bit doubles yield different 16-character strings. This ensures that 
even small floating-point differences can be detected, which may arise 
from compiler implementation differences, precision trade-offs, or 
specific optimizations. While not all inconsistencies indicate compiler 
bugs, detecting them is still valuable for understanding numerical 
instability. If a program exhibits at least one such inconsistency across 
any compiler pair, it is added to the successful set. These programs are 
then reused by the Feedback-Based Mutation strategy to generate programs 
that are more likely to trigger inconsistencies in future iterations.

\section{Experimental Evaluation}
\label{sec:eval}

In this section, we evaluate \vllm to answer the following
research questions:

\begin{itemize}
\item[\textbf{RQ1}] How does \vllm compare to our baselines in terms
of effectiveness at generating floating-point programs that trigger
numerical inconsistencies?

\item[\textbf{RQ2}] What kinds of floating-point inconsistencies is
\vllm most and least effective to trigger?

\item[\textbf{RQ3}] How frequent are numerical inconsistencies between
host and device compilers, or between different host compilers?

\item[\textbf{RQ4}] To what extent do different optimization levels
within the same compiler introduce numerical inconsistencies when
compared to the least intrusive one, i.e., \texttt{-O0\_nofma}?

\end{itemize}

\subsection{Experimental Setup}
\label{subsec:setup}

\subsubsection{Compilers} 

We consider the compilers \gcc version 9.4, \clang version 12.0, and
\nvcc version 12.3. Both \gcc and \clang are used as host compilers to
compile CPU-targeted code, while \nvcc is used as the device compiler
for generating GPU-targeted code only. By default, all host
compilations are linked against the GNU C Library's Math library,
while the device compiler links to the CUDA Math library. We do not
include the \texttt{xlc} compiler (included in the evaluation of
\vari) because we do not have access to machines with the AIX
operating system. However, this evaluation can be easily extended to
other compilers and compiler versions.

\subsubsection{Optimization Levels} 

To align with \vari, we use six optimization levels across all
compilers, listed in \cref{tab:opt_levels}. These levels are designed
to explore the impact of optimization aggressiveness and
floating-point behavior triggering numerical inconsistencies.
\texttt{O0\_nofma} minimizes compiler-induced floating-point
transformations and is the most compliant with the IEEE 754
standard~\cite{DBLP:conf/dagstuhl/Markstein08}. It uses \texttt{-O0}
and explicitly disables Fused Multiply-Add (FMA), which often
introduces more floating-point inconsistencies
\cite{DBLP:conf/ipps/Laguna20}. \texttt{O0} uses \texttt{-O0} with FMA
enabled. \texttt{O1}, \texttt{O2} and \texttt{O3} use increasing
levels of compiler optimizations. \texttt{O3\_fastmath} uses
\texttt{-O3} with aggressive floating-point optimizations, i.e.,
\texttt{-ffast-math} in \gcc/\clang and \texttt{--use\_fast\_math} in
\nvcc, which may break IEEE 754 compliance. \texttt{O3\_fastmath}
provides the highest performance gain, but with the lowest numerical
compliance. We include this level to explore the effect of extreme
optimizations on floating-point consistency, providing a contrast to
more conservative levels.

\begin{table}[t]
\centering
\small 
\caption{Optimization Levels and Compiler Flags}
\label{tab:opt_levels}
\begin{tabular}{lrr}
\toprule
Level & \gcc/\clang & \nvcc \\
\midrule 
\texttt{O0\_nofma}    & \texttt{-O0 -ffp-contract=off}    & \texttt{-O0 --fmad=false}     \\
\texttt{O0}           & \texttt{-O0}                      & \texttt{-O0}                  \\
\texttt{O1}           & \texttt{-O1}                      & \texttt{-O1}                  \\
\texttt{O2}           & \texttt{-O2}                      & \texttt{-O2}                  \\
\texttt{O3}           & \texttt{-O3}                      & \texttt{-O3}                  \\
\texttt{O3\_fastmath} & \texttt{-O3 -ffast-math}          & \texttt{-O3 --use\_fast\_math}\\
\bottomrule
\end{tabular}
\end{table}

\subsubsection{Program Generation Budget and FP Precision} 

For each approach, we set a budget of 1,000 programs, each paired with
a unique set of input values. Every program is compiled using
supported combinations of compilers and optimization levels, and all
resulting binaries are executed. Thus, for each approach there are a
total of 18,000 program runs: 1,000 programs $\times$ 3 compiler pairs
$\times$ 6 optimization levels. By default, the programs use
double-precision floating-point arithmetic, i.e., FP64, but they could
be easily extended to other precisions such as single-precision, i.e.,
FP32.

\subsubsection{Implementation Details}

\vllm is implemented in Python, and uses GPT-4~\cite{DBLP:journals/corr/abs-2303-08774} in the feedback loop to generate floating-point programs since this model is the state-of-the-art for a wide range of NLP-based reasoning tasks \cite{DBLP:journals/corr/abs-2303-12712}. Specifically, we use the \texttt{gpt-4.1-2025-04-14} version of the model. We set \textit{temperature} to 1.2 to increase randomness in token sampling and obtain more varied outputs.
To reduce repetition, we set \textit{frequency\_penalty} to 0.5 to
discourage the model from repeating frequently used tokens, and
\textit{presence\_penalty} to 0.6 to encourage new patterns in the
generated code. These values are informed by findings from prior work
that systematically studies hyperparameter impacts on code generation
quality and diversity~\cite{DBLP:journals/corr/abs-2408-10577}. In addition, each new program is generated using either the Grammar-Based Generation or the Feedback-Based Mutation strategy, with probabilities of 0.3 and 0.7, respectively.

\subsubsection{Environment Settings}
We perform all experiments on a local machine equipped with
an NVIDIA RTX A6000 GPU. The system uses the x86\_64 
architecture and runs Ubuntu 20.04.

\subsection{RQ1: \vllm vs Baselines}
\label{sec:rq1}
The purpose of this RQ is to evaluate the effectiveness of \vllm
compared to various baselines.

\subsubsection{Baselines}

To the best of our knowledge, we are the first to leverage LLMs to
generate floating-point programs that trigger numerical
inconsistencies across compilers and/or optimization levels. To
evaluate the effectiveness of our proposed approach, we consider three
baselines: the \vari framework and two LLM-based variants that differ
in their generation strategy and guidance.

\paragraph{\textbf{\vari~\cite{DBLP:conf/ipps/Laguna20}}} A framework
that triggers floating-point numerical differences in heterogeneous
systems via random program generation. As discussed earlier, \vari
uses a subset of the C language grammar to generate well-formed C/CUDA
floating-point programs, and performs differential testing to find
numerical inconsistencies across compilers and/or optimization levels.

\paragraph{\textbf{\direct}} We refer to our second baseline as
\direct. \direct is an LLM-based variant where we directly instruct
the LLM to generate valid floating-point C programs from scratch using
a carefully designed prompt. In this variant, we do not provide any
grammar specification or examples.

\paragraph{\textbf{\grammar}} Compared to \direct, our third baseline,
referred to as \grammar, augments the prompt with the grammar
specification described in \cref{sec:program-structure}, providing
structural guidance to the LLM and generating programs in a more
Varity-like style. Based on \grammar, our approach \vllm incorporates
a feedback loop that leverages previously generated programs that
triggered numerical inconsistencies. \vllm is guided to generate new,
diverse programs that are more likely to expose other numerical
inconsistencies based on sample programs that also trigger such
inconsistencies.

\subsubsection{Metrics}
\label{sec:metrics}

We evaluate \vllm and our baselines in three dimensions: effectiveness
in triggering numerical inconsistencies, total time cost, and code
diversity.

\paragraph{\textbf{Inconsistency Rate and Count.}} To measure the
effectiveness of each approach at generating floating-point programs
that expose numerical inconsistencies, we define the inconsistency
rate as the ratio of inconsistent output comparisons to the total
number of comparisons performed across compilers and optimization
levels. Let $N$ be the number of generated programs, $C$ be the number
of compilers, and $O$ be the number of optimization levels. Then the
total number of output comparisons is:
\[
\text{Total \# Comparisons} = \binom{C}{2} \times O \times N
\]

For each program, we compare the outputs from every pair of compilers
under each optimization level. As defined in \cref{subsec:finding-incons},
a comparison is marked \emph{inconsistent} if the outputs differ in 
their bitwise representations, as captured by the hexadecimal encoding 
of the floating-point result. The \emph{inconsistency rate} is then:

\[
\text{Inconsistency Rate} = \frac{\text{\# Inconsistencies}}{\text{Total \# Comparisons}}
\]

\paragraph{\textbf{Time Cost.}} We measure the total time taken by
each approach to generate all \( N \) programs, compile each program
with multiple compilers and optimization levels, and execute each
compiled binary with test inputs. This is meant to capture the overall
efficiency of each approach at uncovering numerical inconsistencies at
scale.

\paragraph{\textbf{Program Diversity.}} To evaluate the diversity of
all generated programs, we use two techniques: similarity
scoring~\cite{DBLP:journals/corr/abs-2009-10297} and code clone
detection~\cite{DBLP:journals/access/AinBAAM19}. Measuring diversity
is important to ensure that the LLM is producing structurally and
semantically varied test programs, increasing the chance of exposing a
wider range of floating-point inconsistencies.

We first compute the \codebleu
score~\cite{DBLP:journals/corr/abs-2009-10297}, a widely-used metric
for measuring code similarity. \codebleu absorbs the strength of the
BLEU score~\cite{DBLP:conf/acl/PapineniRWZ02} in the n-gram match, and
further incorporates code syntax via abstract syntax trees and code
semantics via data-flow. A \emph{lower} score indicates more syntactic
and semantic variation, i.e., less similarity between the generated
programs. Specifically, we calculate the pairwise \codebleu score
between all \( N \) generated programs and report the average. 

In addition to \codebleu, we analyze the programs using the NiCad
Clone Detector~\cite{DBLP:conf/iwpc/CordyR11a} to detect redundant
generation patterns that are structurally identical but superficially
different, and thus unlike to reveal different numerical
inconsistencies. NiCad supports the commonly used four types of code
clones, Type-1 through Type-4, as originally proposed by
\cite{roy2007survey}. In our analysis, we focus on Type-1, Type-2, and
Type-2c code clones. Type-1 clones are identical code fragments except
for differences in whitespace or comments. Type-2 clones allow
syntactic similarity with variations in identifiers, literals, and
types. Type-2c, a stricter subtype specific to NiCad, corresponds to
structurally identical code with only consistent renaming of
identifiers. We do not consider Type-3 (clones with inserted, deleted,
or reordered statements), or Type-4 (semantic clones with similar
behavior but different syntax), as floating-point programs are often
computation-intensive and highly sensitive to small syntactic or
semantic changes. Small modifications may lead to significantly
different numerical behaviors, making such code fragments of interest
for potentially triggering numerical inconsistencies.

It is important to acknowledge that even programs not identified as 
clones may still expose the same numerical inconsistencies. To detect 
such cases, a more detailed analysis is necessary. Tools like 
pLiner~\cite{DBLP:conf/sc/0007LR20} or Ciel~\cite{DBLP:conf/supercomputer/MiaoLR23} 
can be employed, since given a program that triggers a numerical 
inconsistency across different compilers or optimization levels, 
these tools pinpoint the specific lines or expressions responsible 
for the inconsistency. Using this information, programs can be 
grouped into equivalence classes based on the root causes of their 
numerical inconsistencies. 

\begin{table}[t]
\centering
\small
\caption{Comparing \vllm with baselines in terms of numerical 
inconsistencies, time cost (hh:mm:ss), and program diversity. Incons. Rate = inconsistency rate; \# Incons. = total number of inconsistencies. Lower \codebleu score is better. }
\label{tab:baselines}
\scalebox{0.95}{
\begin{tabular}{lrrrr}
\toprule
Approach & Incons. Rate & \# Incons. & Time Cost & \codebleu \\
\midrule 
\vari              & 11.93\%    & 2,147       & \textbf{00:30:42}  & 0.3581 \\
\direct            & 14.23\%    & 2,562       & 04:16:37   & 0.3561\\
\grammar           & 16.47\%    & 2,964       & 05:56:13   & 0.3442  \\
\vllm              & \textbf{29.33\%}    & \textbf{5,280}       & 05:37:42  & \textbf{0.2788}\\
\bottomrule
\end{tabular}
}
\end{table}

\subsubsection{Evaluation Results}

\cref{tab:baselines} compares the effectiveness of \vllm and the
baselines at generating floating-point programs that trigger numerical
inconsistencies in terms of the metrics described in
\cref{sec:metrics}. Among all approaches, \vllm achieves the highest
inconsistency rate at 29.33\%, roughly 2.5$\times$ that of \vari,
11.93\%. In total, \vllm discovers 5,280 inconsistencies among the
total 18,000 comparisons across compilers and optimization levels.
This indicates that incorporating the program structural guidance via
the \vari grammar specification and a feedback loop via examples of
inconsistent programs enables the LLM to generate test programs that
are more likely to trigger numerical inconsistencies across compilers
and/or optimization levels. \grammar also outperforms \vari and
\direct by 4.54\% and 2.24\%, respectively. This suggests that
grammar-guided generation combined with the LLM alone improves
inconsistency detection.

In terms of the time cost, \vari remains the most efficient approach
requiring only about 30 minutes to complete the full pipeline for
1,000 generated programs, while LLM-based approaches take
approximately from 4 to 6 hours. The additional cost arises from the
latency of invoking the LLM API during program generation, which
accounts for more than half of the total time. In the future, this
overhead could potentially be reduced by batching API calls or
leveraging local LLM inference to lower latency. Though \vllm has a
longer runtime than \vari, its improvement in inconsistency detection
indicates a trade-off between testing effectiveness and time cost.

Because it is important to avoid repeatedly triggering the same
numerical inconsistencies, diversity in program generation is crucial.
\cref{tab:baselines} compares \vllm with the baselines in terms of the
\codebleu score among the generated programs. A lower score is better,
as it indicates more diverse programs. The results show a clear
descending trend in \codebleu scores from \vari (0.3581) to \direct
(0.3561), \grammar (0.3442), and finally to \vllm (0.2788), suggesting
that program diversity consistently improves across these
approaches. Among them, \vllm achieves the highest diversity, reducing
the \codebleu score by 22.14\% compared to \vari. We attribute this
improvement to the feedback-driven generation mechanism of \vllm,
which iteratively mutates one existing example based on the model's
prior successes. This feedback loop encourages the LLM to explore new
program variants that differ behaviorally from previous ones while
retaining effective structures for triggering numerical
inconsistencies. By explicitly guiding the model to refine and
diversify rather than simply regenerate, \vllm improves the diversity
of program generation while maintaining meaningful program structures.

Regarding Type-1, Type-2, and Type-2c clones, none were detected for any approach, showing that the generated programs across all methods are structurally distinct. This confirms that the prompt design in our LLM-based approaches effectively avoids trivial code duplication while maximizing the discovery of unique inconsistencies.

\begin{mybox}{}

  \textbf{Response to RQ1}: \vllm achieves the highest inconsistency
  rate at 29.33\% and the best program diversity with the lowest
  \codebleu score of 0.2788. This comes to a higher cost due to the
  latency of API LLM calls. On the other hand, \grammar and \direct's
  inconsistency rates are 16.47\% and 14.23\%, while having similar
  time cost as \vllm. Finally, \vari achieves the lowest inconsistency
  rate of 11.93\% but also remains the most efficient approach in
  terms of time cost.

\end{mybox}
\subsection{RQ2: Characterizing Floating-Point Inconsistencies}
\label{sec:rq2}

\subsubsection{Experimental Setup}

Since \vllm outperforms the other LLM-based variants, we focus the
rest of the evaluation on comparing \vllm with \vari with respect to
the floating-point inconsistencies triggered by their generated
floating-point programs. In this RQ, we discuss how often different
classes of floating-point inconsistencies are triggered. We follow
\vari to investigate five classes of floating-point inconsistencies:
\real (normal and subnormal numbers), \zero (positive zero and
negative zero), \posinf (positive infinity), \neginf (negative
infinity), and \nan (not-a-number). Specifically, we count the number
of times that there is a floating-point inconsistency of the kind
$\{\textit{$r_i$}, \textit{$r_j$}\}$, where \textit{r}$_i \neq$
\textit{r}$_j$ and both results belong to different numerical
categories in $\{\real, \zero, \posinf, \neginf, \nan\}$. For example,
an inconsistency between a real number and a zero is counted once as a
\{\real, \zero\} inconsistency.

\subsubsection{Evaluation Results}

\begin{figure}[t]
    \includegraphics[width=\linewidth]{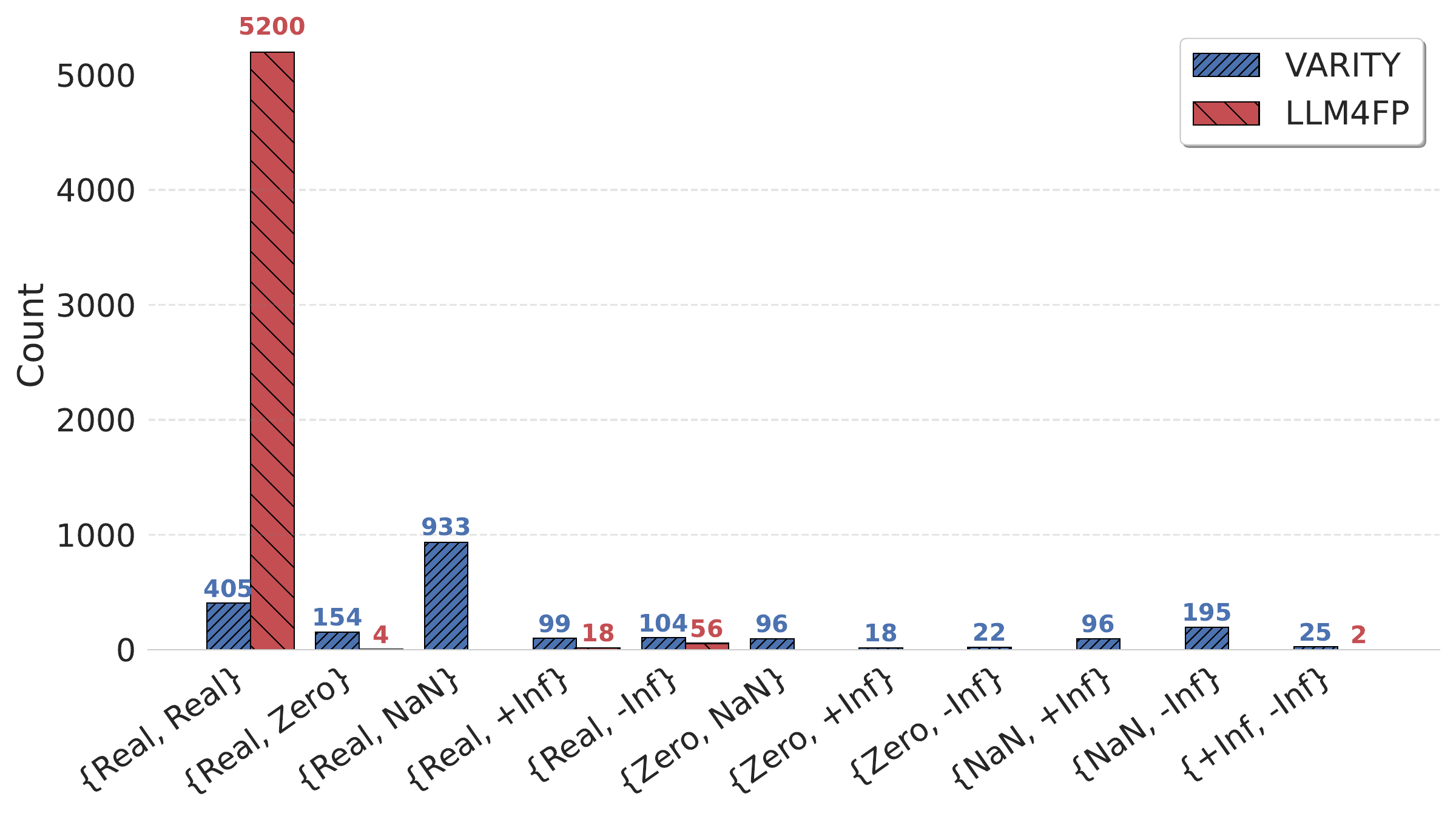}
    \caption{Inconsistency counts of different kinds 
    between two compilers (\vari vs. \vllm)}
    \label{fig:incon_count}
\end{figure}

\begin{table}[t]
\centering
\caption{Inconsistency counts for \vllm across optimization 
levels (``--'' indicates the category did not appear)}
\label{tab:incon_count}
\small
\scalebox{0.9}{
\begin{tabular}{lrrrrr}
\toprule
Level & \real, \real & \real, \zero & \real, \posinf & \real, \neginf & \posinf, \neginf \\
\midrule
\texttt{O0\_nofma}    & 720 & --  & -- & -- & --  \\
\texttt{O0}           & 840 & --  & -- & 4 & --  \\
\texttt{O1}           & 830 & --  & -- & 4 & -- \\
\texttt{O2}           & 814 & --  & -- & 4 & -- \\
\texttt{O3}           & 798 & --  & -- & 4 & -- \\
\texttt{O3\_fastmath} & \textbf{1198} & \textbf{4} & \textbf{18}  & \textbf{40} & \textbf{2}  \\
\midrule
\textbf{Total} &  &  &  &  & \textbf{5,280} \\
\bottomrule
\end{tabular}
}
\end{table}

\begin{table*}[t]
\centering
\caption{Inconsistency rates and digit differences (min / max / avg) across compiler pairs 
at each optimization level for \vari and \vllm (higher inconsistency rates are preferred)}
\label{tab:compiler_pairs_both}
\small
\begin{tabular}{lrrrrrrr}
\toprule
& \multicolumn{3}{c}{\vari} && \multicolumn{3}{c}{\vllm} \\
\cmidrule(lr){2-4} \cmidrule(lr){6-8}
Level & \gcc, \clang & \gcc, \nvcc & \clang, \nvcc && \gcc, \clang & \gcc, \nvcc & \clang, \nvcc \\
\midrule
\texttt{O0\_nofma}    
& \begin{tabular}[t]{r}0.03\%\\($1/1/1.00$)\end{tabular} 
& \begin{tabular}[t]{r}0.67\%\\($1/14/4.73$)\end{tabular} 
& \begin{tabular}[t]{r}0.66\%\\($1/14/4.80$)\end{tabular} 
&& \begin{tabular}[t]{r}0.48\%\\($1/1/1.00$)\end{tabular} 
& \begin{tabular}[t]{r}1.77\%\\($1/15/1.22$)\end{tabular} 
& \begin{tabular}[t]{r}1.75\%\\($1/15/1.23$)\end{tabular} \\
\texttt{O0}           
& \begin{tabular}[t]{r}0.03\%\\($1/1/1.00$)\end{tabular} 
& \begin{tabular}[t]{r}0.69\%\\($1/14/4.58$)\end{tabular} 
& \begin{tabular}[t]{r}0.68\%\\($1/14/4.64$)\end{tabular} 
&& \begin{tabular}[t]{r}0.48\%\\($1/1/1.00$)\end{tabular} 
& \begin{tabular}[t]{r}2.12\%\\($1/15/1.25$)\end{tabular}  
& \begin{tabular}[t]{r}2.09\%\\($1/15/1.25$)\end{tabular} \\
\texttt{O1}           
& \begin{tabular}[t]{r}0.03\%\\($1/1/1.00$)\end{tabular} 
& \begin{tabular}[t]{r}0.70\%\\($1/14/4.56$)\end{tabular} 
& \begin{tabular}[t]{r}0.70\%\\($1/14/4.56$)\end{tabular} 
&& \begin{tabular}[t]{r}0.46\%\\($1/1/1.00$)\end{tabular} 
& \begin{tabular}[t]{r}2.19\%\\($1/15/1.24$)\end{tabular} 
& \begin{tabular}[t]{r}1.98\%\\($1/15/1.27$)\end{tabular} \\
\texttt{O2}           
& \begin{tabular}[t]{r}0.03\%\\($1/1/1.00$)\end{tabular} 
& \begin{tabular}[t]{r}0.70\%\\($1/14/4.56$)\end{tabular} 
& \begin{tabular}[t]{r}0.70\%\\($1/14/4.56$)\end{tabular} 
&& \begin{tabular}[t]{r}0.39\%\\($1/1/1.00$)\end{tabular} 
& \begin{tabular}[t]{r}2.17\%\\($1/15/1.24$)\end{tabular} 
& \begin{tabular}[t]{r}1.98\%\\($1/15/1.27$)\end{tabular} \\
\texttt{O3}           
& \begin{tabular}[t]{r}0.03\%\\($1/1/1.00$)\end{tabular} 
& \begin{tabular}[t]{r}0.70\%\\($1/14/4.56$)\end{tabular} 
& \begin{tabular}[t]{r}0.70\%\\($1/14/4.56$)\end{tabular} 
&& \begin{tabular}[t]{r}0.38\%\\($1/1/1.00$)\end{tabular} 
& \begin{tabular}[t]{r}2.12\%\\($1/15/1.25$)\end{tabular} 
& \begin{tabular}[t]{r}1.95\%\\($1/15/1.27$)\end{tabular} \\
\texttt{O3\_fastmath} 
& \begin{tabular}[t]{r}\textbf{1.02\%}\\($1/16/8.78$)\end{tabular} 
& \begin{tabular}[t]{r}\textbf{1.80\%}\\($1/16/6.75$)\end{tabular} 
& \begin{tabular}[t]{r}\textbf{2.06\%}\\($1/16/7.30$)\end{tabular} 
&& \begin{tabular}[t]{r}\textbf{1.55\%}\\($1/16/3.04$)\end{tabular} 
& \begin{tabular}[t]{r}\textbf{2.64\%}\\($1/16/2.28$)\end{tabular} 
& \begin{tabular}[t]{r}\textbf{2.82}\%\\($1/16/3.08$)\end{tabular} \\
\midrule
Total                & 1.17\% & 5.26\% & \textbf{5.50\%} && 3.74\% & \textbf{13.01\%} & 12.57\% \\
\bottomrule
\end{tabular}
\end{table*}

\cref{fig:incon_count} shows a marked difference between \vllm and
\vari in the distribution of the classes of inconsistencies. \vari's
inconsistencies occur across all kinds, but the majority involve
extreme values like \nan and infinities. On the other hand, 98.48\% of
\vllm's triggered inconsistencies are $\{\real, \real\}$, which are
nearly 13$\times$ more than \vari's. Typically, $\{\real, \real\}$
inconsistencies are specially meaningful since they are much harder
for developers to detect and fix. \vari triggers many extreme values
because it relies on unguided random program generation, which can
produce operations that lead to obvious edge cases, such as division
by zero, or invalid arithmetic.  In contrast, the advantage of using
the LLM is that it tends to generate more structured and semantically
plausible computations that are less likely to cause such catastrophic
outcomes. Instead, \vllm mainly triggers inconsistencies of the
$\{\real, \real\}$ class, which indicate subtle numerical divergence
in normal floating-point computations.

When further breaking down each inconsistency kind by optimization
level for \vllm, we find that the $\{\real, \real\}$ inconsistency
class is observed across all optimization levels, as shown in
\cref{tab:incon_count}. Among them, \texttt{O3\_fastmath} contributes
the largest number of inconsistencies. The extreme values from
$\{\real, \posinf\}$, $\{\real, \neginf\}$, and $\{\neginf, \posinf\}$
are relatively rare and mainly occur under \texttt{O3\_fastmath}. The
results show that the number of inconsistencies across optimization
levels is relatively stable, except for \texttt{O3\_fastmath}, which
consistently amplifies inconsistencies.

\begin{mybox}{}

  \textbf{Response to RQ2}: The most common floating-point
  inconsistency triggered by the programs generated by \vllm is the
  $\{\real, \real\}$ kind, which accounts for 98.48\% of all
  inconsistencies and is observed across all optimization levels. In
  contrast, extreme-value categories involving \posinf, \neginf, or
  \nan are rare and occur in the most aggressive optimization level,
  \texttt{O3\_fastmath}.

\end{mybox}
\subsection{RQ3: Inconsistencies Across Compilers}
\label{sec:rq3}

In this RQ, we analyze whether numerical inconsistencies are more
likely to occur between different host compilers or between host and
device compilers, and how optimization levels influence these
inconsistencies. We also analyze all possible combinations of
compilers to see which combinations produce the largest and smallest
differences. We consider the 16 hexadecimal digits of the floating-point result, and report an inconsistency if any of them do not match. \cref{tab:compiler_pairs_both} summarizes the inconsistency
rates for \vllm and \vari across compiler pairs and optimization
levels. In addition to the rates, it also reports the minimum,
maximum, and average number of differing digits observed in
inconsistent results.

For both approaches, the highest inconsistency rates consistently happen in comparisons between host and device compilers, i.e., \gcc vs. \nvcc and \clang vs. \nvcc, rather than between host compilers such as \gcc vs. \clang. For example, in \vllm, the most inconsistent pair is \gcc vs. \nvcc, with a total inconsistency rate of 13.01\%, compared to 3.74\% for \gcc vs. \clang. A similar trend is observed in \vari, where the highest rate of 5.50\% is from \clang vs. \nvcc, much higher than that of \gcc vs. \clang which is 1.17\%.

Optimization level plays an important role in when and how frequent numerical inconsistencies occur. While similar trends were observed in ~\cref{sec:rq2}, here we focus specifically on inconsistencies between host and device compilers. For \vari, the highest inconsistency rates consistently occur at the \texttt{O3\_fastmath} level, reaching up to 2.06\% for \clang vs. \nvcc, while the rates at other optimization levels remain very low, mostly below 1\%. This is expected as aggressive optimizations and fast-math transformations can increase the chance of observing inconsistencies across compilers. For \vllm, the \texttt{O3\_fastmath} level still yields the highest inconsistency rates across compiler pairs, peaking at 2.82\% for \clang vs. \nvcc. However, unlike \vari, other optimization levels also exhibit relatively high rates around 2\%, suggesting that \vllm triggers inconsistencies more broadly across compiler settings. This observation highlights the potential of \vllm for generating more complex and diverse floating-point programs that expose subtle compiler differences beyond aggressive optimization scenarios.

Beyond inconsistency rates, the digit difference results provide additional insight into the severity of numerical discrepancies. On average, \vllm inconsistencies exhibit smaller digit differences, i.e., from around 1 to 3 digits, compared to \vari, whose average differences are typically between 4 and 8 digits. This suggests that while \vllm exposes inconsistencies more frequently, most of them correspond to subtle numerical divergences rather than large-magnitude errors. In particular, \vllm mainly triggers inconsistencies of the $\{\real, \real\}$ class, reflecting sensitive variations in normal floating-point computations rather than catastrophic precision loss.

\begin{mybox}{}

  \textbf{Response to RQ3}: Host–device numerical inconsistencies are more frequent than host–host ones. Compared to \vari, \vllm discovers more inconsistencies even at lower optimization levels, indicating its ability to expose numerical differences beyond those revealed by aggressive optimizations. \vllm also exposes smaller digit differences, indicating that most numerical discrepancies are subtle rather than large-magnitude errors.
\end{mybox}
\subsection{RQ4: Comparing to \texttt{O0} without FMA}
\label{sec:rq4}

\begin{table}[t]
\small
\centering
\caption{Inconsistency rates between any optimization 
level and \texttt{O0\_nofma} for \vari and \vllm (higher 
inconsistency rates are preferred, and ``--'' indicates there is no inconsistency)}
\label{tab:opt_pairs_both}

\begin{tabular}{lrrrrrrr}
\toprule
& \multicolumn{3}{c}{\vari} && \multicolumn{3}{c}{\vllm} \\
\cmidrule(lr){2-4}\cmidrule(lr){6-8}
Level & \gcc  & \clang  & \nvcc && \gcc & \clang & \nvcc\\
\midrule
\texttt{O0}           & -- & --  & 0.03\%&& -- & -- & 0.51\%  \\
\texttt{O1}           & 0.02\% & 0.04\% & 0.03\%&& 0.36\% & 0.75\% & 0.51\% \\
\texttt{O2}           & 0.02\% & 0.04\% & 0.03\%&& 0.41\% & 0.77\% & 0.51\% \\
\texttt{O3}           & 0.01\% & 0.04\% & 0.03\%&& 0.38\% & 0.78\% & 0.51\% \\
\texttt{O3\_fastmath} & \textbf{1.61\%} & \textbf{2.07\%}  &
\textbf{0.04\%}&& \textbf{1.76\%} & \textbf{2.30\%} & \textbf{0.52\%} \\
\midrule
Total       & 1.66\% & \textbf{2.19\%} & 0.16\% && 2.91\% &
\textbf{4.60\%} & 2.56\%\\
\bottomrule
\end{tabular}

\end{table}

This section compares the results of \texttt{-O0\_nofma} to all other
optimization levels. Since \texttt{-O0\_nofma} is the level most
compliant with IEEE, we aim to measure how much other optimization
levels differ from it within each compiler. \cref{tab:opt_pairs_both} shows
the percentage of times that there is a difference between
\texttt{-O0\_nofma} and each other optimization level for \vari and
\vllm. All comparisons are done within the same compiler across
different optimization levels.

As shown in \cref{tab:opt_pairs_both}, the inconsistency rate varies
when comparing \texttt{O0\_nofma} with different optimization
levels. For \vari, the inconsistency rates at lower optimization
levels from \texttt{O0} to \texttt{O3} remain near zero, all below
0.05\%. It mainly detects differences caused by the most aggressive
optimization level, \texttt{O3\_fastmath}, especially for \clang which
shows an inconsistency rate of 2.07\% compared to \texttt{O0\_nofma}.
In contrast, \vllm reports higher rates across all levels, with
noticeable differences at moderate optimizations, reaching up to
0.78\% for \clang. The results indicate that \vllm's programs are
better at finding variations even when optimizations are less
aggressive. For both \vari and \vllm, \clang has the highest number of
floating-point inconsistencies across its optimization levels, while
\nvcc is the most stable. Compared to \vari, \vllm finds more
variations across optimization levels within each compiler.

\begin{mybox}{}

  \textbf{Response to RQ4}: Different optimizations within the same
  compiler can introduce different floating-point inconsistencies compared to \texttt{-O0\_nofma}. \vllm finds inconsistencies across all optimization levels, not just the most aggressive ones. \clang has the most inconsistencies while \nvcc is the most stable.

\end{mybox}

\section{Limitations}
\label{sec:limitations}

While \vllm improves the state of the art in generating floating-point
programs that trigger numerical inconsistencies, it also has several
limitations. First, our approach has a
higher runtime cost due to the latency of LLM API calls which may
limit scalability for large-scale testing. Nevertheless, the improved
effectiveness of \vllm helps offset this overhead. Second, while we
explicitly instruct the LLM to avoid undefined behavior, we cannot
guarantee its absence, which may cause false positives in
inconsistency detection. In the future, we plan to explore
constraining our program generation rules and integrating runtime
undefined behavior checkers such as GCC/Clang Undefined Behavior
Sanitizer or AddressSanitizer~\cite{DBLP:conf/usenix/SerebryanyBPV12}
to reduce false positives while maintaining valid program semantics.

Another limitation concerns prompt design and the interpretation of
detected floating-point inconsistencies. The effectiveness of \vllm
depends heavily on carefully crafted prompts. In practice, we find
that good prompts combine structural and semantic constraints with
explicit formatting rules and targeted mutation strategies, often
supported by a successful example, whereas poor prompts tend to be
vague or unconstrained, which can result in trivial or uninformative
programs. If the prompts are poorly designed, the effectiveness of
triggering numerical inconsistencies will be limited. Lastly, not all
inconsistencies detected by \vllm indicate real bugs. Some may reflect
allowed differences in compiler behavior, while others can impact
program stability and lead to incorrect results. In either case, prior
work~\cite{DBLP:conf/ipps/Laguna20, DBLP:conf/iiswc/SawayaBBGA17,
DBLP:conf/sc/0007LR20, DBLP:conf/supercomputer/MiaoLR23,
DBLP:conf/ics/MiaoLR24} has shown that it is important to be aware of
floating-point inconsistencies across compiler and/or compiler
optimizations.

\section{Related Work}
\label{sec:related}

Compiler testing has been widely studied to ensure the correctness and
reliability of compiler
behavior~\cite{DBLP:journals/csur/ChenPPXZHZ20}. Traditional compiler
testing techniques include random
testing~\cite{DBLP:journals/tse/DuranN84, DBLP:journals/cj/InceH86},
randomized differential testing~\cite{DBLP:journals/dtj/McKeeman98,
  DBLP:conf/icse/GroceHJ07}, equivalence modulo inputs
testing~\cite{DBLP:conf/pldi/LeAS14}, program
enumeration~\cite{DBLP:conf/pldi/ZhangSS17} among others. The main
challenge in compiler testing is to generate diverse, syntactically
and semantically valid programs to test the compiler
behaviors. Grammar-based approaches synthesize programs from scratch
by following generation rules to ensure syntactic validity, such as
CSmith~\cite{DBLP:conf/pldi/YangCER11},
YARPGen~\cite{DBLP:journals/pacmpl/LivinskiiBR20}, and
LangFuzz~\cite{DBLP:conf/uss/HollerHZ12}. To generate more realistic
and complicated test programs, mutation-based
approaches~\cite{DBLP:journals/pacmpl/DonaldsonELT17,
  DBLP:conf/pldi/LeAS14, DBLP:conf/oopsla/LeSS15,
  DBLP:conf/pldi/ZhangSS17} generate new programs by mutating valid
seed programs.

For floating-point programs, FLiT~\cite{DBLP:conf/iiswc/SawayaBBGA17}
provides a test infrastructure for detecting variability in
floating-point code across compilers and platforms. Unlike program
generation approaches, FLiT does not synthesize new tests but instead
systematically runs existing workloads to uncover cross-platform
result differences. \vari~\cite{DBLP:conf/ipps/Laguna20} extends
compiler testing to this domain by using grammar-guided random
generation to synthesize test programs, focusing on numerical
inconsistencies across different compilers and architectures rather
than compiler bugs. However, it does not leverage domain knowledge or
feedback from previous tests, which limits its ability to discover
floating-point inconsistencies efficiently. In contrast, our work uses
LLMs to generate and adapt test programs based on prior results,
enabling us to discover more floating-point inconsistencies across
compilers.

Several prior tools address different aspects of compiler-induced
numerical inconsistencies but are complementary to our
program-generation focus. pLiner~\cite{DBLP:conf/sc/0007LR20} isolates
known numerical inconsistencies in C/C++ CPU programs at the line
level using hierarchical code isolation and precision enhancement,
while Ciel~\cite{DBLP:conf/supercomputer/MiaoLR23} extends it to
heterogeneous programs and isolates inconsistencies at the expression
level, enabling finer-grained analysis. Tools such as
CIV~\cite{DBLP:conf/wcre/YuYYLCH23},
S3FP~\cite{DBLP:conf/ppopp/ChiangGRS14}, and
CIGEN~\cite{DBLP:conf/ics/MiaoLR24} are input-generation methods that
aim to expose floating-point inconsistencies or
high-precision errors in existing programs, rather than synthesizing
new programs. FPDiff~\cite{DBLP:conf/issta/VanoverDR20} performs
differential testing on automatically identified synonymous functions
across numerical libraries to detect inconsistencies in their results
under specific inputs. Together, these approaches focus on isolating
or exposing inconsistencies in existing programs or libraries, whereas
our work targets automatic generation of new floating-point test
programs to discover inconsistencies across compilers from scratch.

Recent advances in LLMs have opened new opportunities for test program
generation~\cite{DBLP:journals/corr/abs-2406-00515}. LLMs like
Codex~\cite{DBLP:journals/corr/abs-2107-03374} and
GPT~\cite{DBLP:journals/corr/abs-2303-08774} have demonstrated strong
capabilities in generating syntactically correct and semantically
meaningful code from natural language prompts in tools like
TestPilot~\cite{DBLP:journals/corr/abs-2302-06527} and
ChatTester~\cite{DBLP:journals/corr/abs-2305-04207}. Several recent
works have applied LLMs to testing compilers
\cite{DBLP:conf/icse/XiaPTP024, DBLP:journals/pacmpl/YangDLY0J024,
DBLP:journals/fgcs/MunleyJC24, DBLP:conf/sigsoft/Gu23}. For example,
WhiteFox~\cite{DBLP:journals/pacmpl/YangDLY0J024} integrates LLMs with
compiler source code to guide white-box fuzzing for optimization
testing. Fuzz4All~\cite{DBLP:conf/icse/XiaPTP024} uses LLMs to fuzz
compilers across multiple languages via auto-prompting and iterative
mutation. Unlike our work, they focus on general bug detection and do
not target floating-point inconsistencies. Overall, our work bridges
the gap between LLM-based program generation and numerical
inconsistency detection across compilers, which is underexplored in
prior research.

\section{Conclusion}
\label{sec:conclusion}

We propose \vllm, a novel LLM-based approach for generating 
floating-point programs that trigger numerical inconsistencies 
across compilers and/or optimization levels. By combining Grammar-Based 
Generation and Feedback-Based Mutation, \vllm significantly outperforms 
existing tools like \vari, detecting nearly 2.5 times as many inconsistencies. 
Our evaluation shows improved program diversity over other baselines, with a trade-off in time cost. Most inconsistencies it finds 
involve real-valued results, and they demonstrate \vllm's effectiveness at 
exposing inconsistencies across a wide range of optimization levels 
and between host and device compilers. Our code, documentation and experimental data are publicly available at \url{https://github.com/ucd-plse/LLM4FP}.

\begin{acks}
  
This work was supported in part by the National Science Foundation
under award CNS-2346396, and the U.S. Department of Energy, Office of
Science, Advanced Scientific Computing Research, under award
DE-SC0020286.

\end{acks}

\balance
\bibliographystyle{ACM-Reference-Format}
\bibliography{main}

\end{document}